\documentclass[preprint,showpacs,preprintnumbers,amsmath,amssymb]{revtex4}

\usepackage{graphicx}% Include figure files
\usepackage{dcolumn}% Align table columns on decimal point

\begin{document}

\preprint{APS/123-QED}

\title{Electrostatic attraction between cationic-anionic assemblies with surface compositional
heterogeneities}%

\author{Y. S. Velichko}%
%\email{y-velichko@northwestern.edu}%
\author{M. Olvera de la Cruz}%
\email{m-olvera@northwestern.edu}%
\affiliation{Department of Materials Science and Engineering,
Northwestern University, IL 60208, USA.}%

\date{\today}% It is always \today, today,

\begin{abstract}
Electrostatics plays a key role in biomolecular assembly. Oppositely charged
biomolecules, for instance, can co-assembled into functional units, such as
DNA and histone proteins into nucleosomes and actin-binding protein complexes
into cytoskeleton components, at appropriate ionic conditions. These
cationic-anionic co-assemblies often have surface charge heterogeneities that
result from the delicate balance between electrostatics and packing
constraints. Despite their importance, the precise role of surface charge
heterogeneities in the organization of cationic-anionic co-assemblies is not
well understood. We show here that co-assemblies with charge heterogeneities
strongly interact through polarization of the domains. We find that this
leads to symmetry breaking, which is important for functional capabilities,
and structural changes, which is crucial in the organization of
co-assemblies. We determine the range and strength of the attraction as a
function of the competition between the steric and hydrophobic constraints
and electrostatic interactions.
\end{abstract}



\maketitle

\section{Introduction}

The structure of biomolecular assemblies of cationic and anionic molecules
containing hydrophobic groups is controlled by the architecture of the
molecules and the intermolecular interactions of the components. The
co-assemblies can have spherical, cylindrical or planar symmetries.
Oppositely charged lipids, for example, co-assemble into cylindrical
micelles\cite{KalerJCP1992} or into vesicles\cite{ZembNature2001}, while
mixtures of cationic and anionic peptide amphiphiles capable of beta sheet
formation co-assemble into
fibers\cite{StuppScience2001,StuppJACS2003,StuppScience2004}. The stability
of these assemblies is determined by their surface properties, which in turn
is the result of the competition between the different type of interactions,
similar to proteins\cite{Whitesides}. The adsorption of charged lipids onto
oppositely charged surfaces\cite{IsraelachviliPNAS2005} is a simple example
of planar co-assembly with competing interactions where electrostatics favors
homogeneous surface coverage and the lipids hydrophobic tails favor bilayer
formation with local excess charge leading to surfaces with charge
inhomogeneities. These charge heterogeneities are expected in many
cationic-anionic co-assemblies including actin and microtubular associating
proteins (MAP) complexes\cite{Actin2}, as well as in nucleosomes. Herein we
analyze interactions among co-assemblies with surface charge heterogeneities.
We show that co-assemblies with homogeneous surfaces interact weakly and
induce small changes in the surface heterogeneities of neighboring assemblies
suggesting that these co-assemblies have decreased functionality. Meanwhile,
co-assemblies with large surface heterogeneities interact strongly and are
capable of large structural modifications.

What controls the surface heterogeneities in cationic-anionic co-assemblies?
The net incompatibility among chemically different charged components, which
in water can be due to different degrees of hydrophobicity, promotes
macroscopic segregation of the different components. Meanwhile,
electrostatics favors correlated ionic crystal structures. The competition of
these interactions result in the formation of surface charge
patterns\cite{IsraelachviliPNAS2005,SolisJCP2005,VelichkoPRE2005,LoverdeJCP2006}
in aqueous interfaces due to the high permittivity of the water. In dense
media charges are paired, while in water they can be dissociated. At the
interface between a dense media and an aqueous solution, charged patterns of
different length scales are possible because the mean permittivity of the two
media is sufficiently high. The formation of large domains, which decrease
the interface among the components is restricted due to the penalty
associated with charge accumulation. Surprisingly this penalty may be
overcome in the presence of other assemblies, by correlating the oppositely
charged domains on the neighboring assemblies. Our computer simulations and
theoretical analysis show that the attraction mechanism among co-assemblies
is strongly dependent on the degree of compatibility among the co-assembled
molecules. Surfaces of co-assembled molecules with large degrees of
incompatibility have strong attractions which decay exponentially with
separation distance, and the range and strength grow as the degree of
incompatibility among the components increases. Compatible co-assembled
components have both weak correlated interactions at short distances, and
weak dispersion type forces at longer distances similar to the attraction
among charged surfaces via their counterions\cite{Ninham,Stevens,Marcelja}.

Formation of surface patterns is not only the result of the interplay of
short-- and long--range forces. The geometry of the assembly also plays a key
role. Cylindrical geometry, where the charged units are confined to the
surface of a cylinder, is common in biomolecular assemblies. The appearance
of large domains on the surface of cylindrical co-assemblies induce
restrictions on the size and symmetry of the domains that affect their
physical properties. Such assemblies have attracted much attention as they
provide a possible way to create tunable functional biomolecular materials,
such as stable vesicles for drug delivery~\cite{ZembNature2001,KalerJCP1992}
and bio--active fibers~\cite{StuppJACS2003,StuppScience2004}.

Bundle formation is highly prominent in systems of fibers. Even weak
dispersion forces lead to aggregation. In many biological systems the
attractions, however, are strong and specific. Here we show that
electrostatic interactions are important by modelling cylindrical assemblies
of molecules with chemically incompatible oppositely charged head groups
exposed onto the surface. This is a coarse grained model for cationic-anionic
amphiphile co-assemblies or for self-attracting (hydrophobic) charged
particles adsorbed on the surface of oppositely charged fibers such as
actin-MAP complexes where the charged proteins effectively change the local
properties including local charge density and degree of
hydrophobicity~\cite{SandersPRL2005}. We consider each aggregate as a stable
structure, thus constraining the cylindrical geometry and allowing molecules
to move only on the surface. The incompatibility among oppositely charged
co-assembled molecules per thermal energy $K_BT$, denoted by $\chi$ (expected
to be a complex function which includes all the terms opposing mixing of
molecules), and the electrostatic interactions determine the surface charge
heterogeneities. The study of these interacting assemblies sheds light on the
mechanism of non--specific interactions in many biological systems with
surface compositional heterogeneities.

The interaction among assemblies with surface compositional heterogeneities
have some analogies with the attraction between charged membranes or
polyelectrolytes in the solution of neutralizing
counterions~\cite{Stevens,Rouzina,BarratACP1996,HaPRE1998,ArenzonEPJB1999,SolisPRE1999,ChargedMembranes,Pincus}.
However, there is a significant difference. In particular, in case of charged
surfaces in the solution of neutralizing counterions, the size of the Wigner
crystal cell is determined by the conditions of surface electro-neutrality
and $\Lambda\propto\sqrt{Q/(e\sigma)}$, where $e\sigma$ is a surface charge
density and $Q$ is the charge of condensed counterions~\cite{Rouzina}. The
fluctuations of condensed ions and structural correlations induce
electrostatic attraction. On the other hand, in case of the surfaces with
compositional heterogeneities, the size of the Wigner crystal cell is
determined by a balance of electrostatic and interfacial energies and
$\Lambda\propto\sqrt{\chi/\sigma^2}$. It should be also noticed that the area
of the domain determines the effective charge $Q\propto L^2$, while charge of
condensed counterion remains the same.

In this paper we study the interaction among cylindrical assemblies with
surface compositional heterogenous. We show that the compositional
heterogeneities of two interacting cylindrical assemblies are correlated and
the size of heterogeneities grows as the assemblies approach each other. This
leads to a strong attraction. We show that the effective potential of
interaction depends on the degree of surface compositional segregation and
the separation distance among assemblies. The range and the strength of the
attraction increase as the degree of the net incompatibility of the
cationic--anionic components increases.

\section{Simulation Method}
We consider each aggregate as a stable structure, thus constraining the
cylindrical geometry and allowing molecules to move only on the surface.
Within the framework of the primitive model, each aggregates is composed of a
stoichiometric mixture of equal number $N_+=N_-$ of positively and negatively
charged units of equal absolute charge $|Q_+|=|Q_-|=1$ and diameter $a=1$.
The cylindrical assemblies are placed into the box $L_x\times L_y\times L_z$
in the middle of the $YZ-$plane along $z-$axis, where $L_y=L_x+2R_c+D$. We
study the interaction of two parallel cylindrical assemblies of radius $R_c$
and length $L_z$ separated by the distance $2R_c+D$, where $D$ gives the
distance between ions on the faced surfaces of the neighboring assemblies.
From here, due to the fixed radius $R_c$, we will use only $D$ to indicate
the separation distance.

The surface of each cylinder is filled with units in such a way that all
units are placed into the knots of a triangular lattice of
period~$a$~(Fig.~\ref{fig-1}) and thus the number surface density is
$\sigma=2/\sqrt{3}a^{2}$. The short-ranged interaction between units is taken
into account via a simplistic discrete van der Waals potential, due to the
discreet model. The net degree of compatibility among different components,
i.e. positively and negatively charged units, is controlled by Flory--Huggins
parameter
$\chi=[\varepsilon_{ij}-\frac{1}{2}(\varepsilon_{ii}+\varepsilon_{jj})]/k_BT$,
where $\varepsilon_{ij}$ represents a pair interaction energy among $i$ and
$j$ ions in $k_BT$ units. Compositional incompatibility results in the
formation of segregated domains, whose growth is inhibited by electrostatics.
We use Debye--Hueckel potential to describe the electrostatic interactions
\begin{eqnarray}
u(r)/k_BT=\ell\frac{e^{-\kappa r}}{r}
\end{eqnarray}
where $\ell=e^2/4\pi\epsilon_{\rm o}k_BT$ is Bjerrum length, $\epsilon_{\rm
o}$ is the average dielectric constant of the media in units $e^2/4\pi a$ and
$\kappa=4\pi\ell\sum c_iz_i^2$ is the inverse Debye screening length, $c_i$
is the concentration and $z_i$ is the valence of the $i-$th neutralizing
counterion in the solution. Taking into account a biological origin of the
system, i.e. the large head groups size $(\sim 10\textrm{\AA})$ and high
average dielectric permittivity of the aqueous solution $(\sim 80)$, the
Bjerrum length $\ell=0.2a$ is considered. Due to the assemblies are
electroneutral, counterion condensation in not expected in the wide range of
temperatures and, thus, we choose Debye length $a/\kappa=25$.

We report standard canonical Monte Carlo simulations following the Metropolis
scheme for various values of $\chi\in[0.5,12]$, $R_{c}/a=2-5$, $L_{x}=L_{z}$
and $L_{z}/a=100$. Simple moves in the phase space are performed by exchange
of two randomly chosen particles. Each system is equilibrated during $10^5$
MC steps per particle and another $10^5$ MC steps are used to perform
measurements. The equilibration process is accompanied by a gradual decrease
of temperature (temperature annealing) from $T_{max}=10$ to $T_{min}=1$.

\section{Results}
The competition between the net short range repulsion among dissimilar
components and the electrostatic penalty associated with the charge
accumulation results in the formation of the surface composition
heterogeneities. With increase in $\chi$ domains size grows, however the
electrostatics restrict the possibility of macroscopic segregation. The
steric commensurability between the surface charge patterns applies
geometrical restrictions on the size and the symmetry of the domains, and
together with the long--range electrostatic interactions induces periodic
arrangement of domains of opposite charge, which stabilizes the co-assembled
structure in analogy with the Wigner crystals of ions~\cite{Madelung}.
Figure~\ref{fig-2}~(a) shows that with increase in $\chi$ the normalized
energy of the system $U/\chi N$ is decreasing for both, nearby $(D/\sigma=1)$
and separated assemblies. The energy of the nearby assemblies is notably
smaller, suggesting a strong attraction among aggregates.

With increase in $\chi$ the system undergos transition from the isotropic to
segregated phase that is indicated as a broad peak on the heat
capacity~(Fig.~\ref{fig-2}~(b)) at $\chi$ denoted by $\chi_1$. For nearby
assemblies the peak is higher and it is shifted to smaller $\chi$ values,
suggesting a promotion of the compositional segregation by the assemblies
cross--interaction.

\subsection{Theoretical model}
Due to complexity of analyzes of cylindrical assemblies, we start by
considering two interacting planar surfaces. In the limit of small  net
incompatibility, the contribution of density fluctuation to the mean free
energy can be analyze by means of Random Phase Approximation
(RPA)~\cite{BorueMacromol1988}. In the limit of small density fluctuations
around an homogenous state of mean density $\bar{\rho}$ equal to the number
fraction of positively and negatively charged components
$\bar{\rho}=\bar{\rho}_+=\bar{\rho}_-$. The total free energy $F$ including
one loop corrections (OLC) can be obtained from the partition function
$Z=e^{-F/k_BT}$. The partition function of the system in terms of the Fourier
components of the density $\rho_q$ reads
\begin{eqnarray}\label{OLCZ}
Z=Z_{\rm o}\int\exp\left(-\frac{1}{2V}\sum\limits_{q\neq 0}
{\rho_q}\mathbb{A}_q{\rho_q^{\textrm{T}}}\right)\prod\limits_{q>0}\frac{d\rho_q}{V}
\end{eqnarray}
where $Z_{\rm o}=e^{-F_{\rm o}/k_{\rm B}T}$ is zero mode partition function
($q=0$) with no fluctuation contribution, ${\rho_q}=(\rho_q^1,\rho_q^2),$
where $\rho_q^1$ and $\rho_q^2$ are the Fourier components of the composition
fluctuations around the mean value, $\Delta\rho^i(r)=\rho^i(r)-\bar{\rho}$,
of cylinder $i=1,2$, respectively. $\mathbb{A}$ is the inverse of the density
correlation matrix $\mathbb{S}$ with elements
$s_{ij}(q)=\left<\rho^{i}_q\rho^{j}_{-q}\right>$. The diagonal elements
$a_{ii}(q)$ of the matrix $\mathbb{A}$ can be estimated according to
Cahn--Hilliard free energy of a neutral system~\cite{CahnJCP1958} with
additional gradient (interfacial energy) and electrostatic terms
\begin{eqnarray}
a_{ii}(q)=\frac{1}{\bar{\rho}(1-\bar{\rho})}-2\chi
+\chi\frac{(qa)}{2}^2+4\sigma u_{ii}(q)
\end{eqnarray}
where the electrostatic potential
\begin{eqnarray}
u_{ii}(q)=\ell\int \frac{e^{\kappa r}}{r} e^{irq}d\mathbf{r}
=\frac{2\pi\ell}{\sqrt{q^2+{\kappa}^2}}.
\end{eqnarray}
On the other hand, non--diagonal terms, $a_{ij}(q,D)= 4\sigma u_{ij}(q,D),$
include only the energy of the electrostatic interaction between the
separated surfaces
\begin{eqnarray}
u_{ij}(q,D)=\ell\int
\frac{e^{\kappa\sqrt{r^2+D^2}}}{\sqrt{r^2+D^2}}e^{irq}d\mathbf{r}\nonumber\\
=\frac{2\pi\ell}{\sqrt{q^2+{\kappa}^2}}e^{-D\sqrt{q^2+{\kappa}^2}},
\end{eqnarray}
where $D$ is the distance between the surfaces.

\subsection{Domains polarization}
In the limit $D\rightarrow\infty$, when assemblies do not interact, the
competition between electrostatics, $4\sigma u_{ii}(q)$, and interfacial,
$\chi(qa)^2/2$, energies result in the formation of favorable density
fluctuations of finite wave length $q^*_o$ and appearance of a peak in
$s_{11}(q)$ at $q^*_o\simeq(8\pi\ell\sigma/\chi
a^2)^{1/3}$(Fig.~\ref{fig-crit}~{(a)}). Besides, detailed analysis of binary
cationic-anionic mixtures restricted on the surface of
cylinders~\cite{VelichkoPRE2005} and planes~\cite{LoverdeJCP2006} reveal many
interesting final temperature effects as well as stripe structures along the
assembly at lower temperatures.

As one cylinder approaches another, long range electrostatics correlates
surface compositional heterogeneities on the neighboring assemblies to
minimize the electrostatic energy. In closed vicinity, $D<R_c$, domains of
opposite charge are located in front of each other and electrostatic energy
of the cross interaction effectively compensates penalty associated with
charge accumulation. That results in the polarization or enlargement of the
domains. The peak position is shifting in direction of small $q$ values as
$D$ decreases (Fig.~\ref{fig-crit}~{(a)}). In the limit $q^*_oD>1$, the
dependence of the domain size on the distance $D$ among assemblies,
calculated form the $\partial s_{11}/\partial q=0,$ has simple approximation
\begin{eqnarray}
q^*\simeq q^*_o(1-q^*_oD e^{-q^*_oD})^{1/3}.
\end{eqnarray}
At the same time, the magnitude of the peak $S(q^*)$ is growing with decrease
in distance that indicates strengthening of the compositional segregation.
Figure~\ref{fig-crit}~{(b)} shows the OLC critical conditions, obtained from
the numerical solution of equation $s^{-1}_{11}(q^*_o)=0$.

To analyze the domains polarization and compare results of the computer
simulation and theory we calculate the static structure factor along the
assemblies faced surface
\begin{equation}
S_z(q)=\frac{1}{N_{+}}\left<\left| \sum\limits_{i,j=0}^{N_{+}}e^{\imath
q\cdot\Delta z_{ij}}\right|^2\right>,
\end{equation}
where $\Delta z_{ij}=z_{i}-z_{j}$ for all ion pairs with equal $X$ and $Y$
components, $x_{i}=x_{j}$ and $y_{i}=y_{j}$, where $\vec{r}_{i}=\left(x_{i},
y_{i}, z_{i}\right)$ is a cartesian vector. The competition between
electrostatic repulsion and interfacial energies results in the formation of
most favorable fluctuations of finite wave length~$\lambda=2\pi/q^*$ and the
appearance of a peak in~$S_z(q)$ at~$q^*$. Figure~\ref{fig-4}~{(b)} shows the
dependence of $q^*$ versus $D$ for different values of $\chi$. With decrease
in the separation distance the $q^*$ is also decreasing, indicating the
domains enlargement. For $\chi<\chi^{(1)},$ the $q^*$ always has a final
value, while for $\chi>\chi^{(1)}$ for nearby assemblies the domains size
becomes in order of the assemblies length (Fig.~\ref{fig-4}~{(a)}) indicating
a very strong segregation. It is important to underline, that in this case
cylindrical geometry results in breaking of symmetry of surface domains, due
to $\Lambda\sim R_{c}$, and then domains grow only along the assembly.
Polarization of domains is not only a new and interesting phenomenon that may
be used to explain aggregation of different assemblies or biomolecules into
bundles. It also affects the effective potential of attraction, due to the
dependence of domain size $\Lambda=2\pi/q^*$ on the separation distance $D$.
Polarization of domains shows another significant difference in the nature of
interaction among assemblies with surface compositional heterogeneities and
charged membranes or polyelectrolytes in the solution of neutralizing
counterions.

\subsection{Electrostatic attraction}
The correlation and polarization of surface domains decreases the energy of
the system, suggesting the attraction among the assemblies. The strength of
the attraction can be calculated from the dependence of energy $U/\chi N$ on
the distance $D$ among assemblies, as $\Delta U(D)/N=(U(D)-U_o)/N$, where
$U_o$ is the energy of two separated assemblies. At the same time, the energy
of two interacting planar assemblies can be analyzed analytically as
\begin{eqnarray}\label{dE2d}
\Delta E_{2d}=-T^2\frac{\partial(F/T)}{\partial T},
\end{eqnarray}
where $F$ is the OLC free energy (Eq.~\ref{OLCZ}). Figures~\ref{fig-5} (a)
and (b) show both of the energies for different values of the net
incompatibility $\chi$.

In the limit of small net incompatibility, $\chi<\chi_o$, when the surfaces
are nearly homogeneous and isotropic, the attraction is originates by
cooperative correlations of charge fluctuations leading to a long-ranged
power law attraction, $D^{-2}$~\cite{Pincus}. With decrease in the distance
$D$, the density fluctuations are suppressed due to the lateral repulsions
among ordered domains. That leads to stronger attraction. On the other hand,
with increase in the degree of net incompatibility $\chi$ the strength of the
attraction is also growing, due to the increasing magnitude of the density
fluctuation. However, as we will show later, in the limit of strong
incompatibility, $\chi>\chi_o$, we find an unexpected slowly decaying
exponential attraction.

In the limit of strong net incompatibility, when surface domains are well
pronounced, charged domains correlate themselves to minimize the energy and
form one dimensional periodic array. Therefore, the ground state
configuration is a pair of two cylindrical assemblies with a charged surface
pattern of positive and negative domains in one assembly in front of an
oppositely charged domains in the second assembly. In that case, the
attraction energy can be approximated as
\begin{eqnarray}\label{ZT1Dim}
\Delta U_{1d}/N =-\frac{(e\sigma R_{c}\Lambda)^2}{\epsilon_o}
\sum\limits_{m=-\infty}^{\infty}\frac{(-1)^m}{\sqrt{D^2+(\Lambda
m)^2}}\nonumber\\
=-\frac{\Lambda(e\sigma R_{c})^2}{2\epsilon_o}\sum\limits_{m=1}^{\infty}
K_o\left[\pi\frac{Dm}{\Lambda}\right]\left\{1-\cos(\pi m)\right\}\\
\simeq-\frac{(e\sigma R_{c})^2\Lambda }{\epsilon_o}
\sqrt{\frac{\Lambda}{2D}}e^{-\pi\frac{D}{\Lambda}}. \nonumber
\end{eqnarray}
This approximation should work in the limit of thin cylinders, when domain
size $\Lambda$ is large then the radius of the assembly $R_{c}$. On the other
hand, when $\Lambda<R_{c}$, cylindrical assemblies can not be considered as
one dimensional objects and a planar surface represents another limit. In
this case, the attraction energy scales as
\begin{eqnarray}\label{ZT2Dim}
\Delta U_{2d}/N \simeq-\frac{(e\sigma)^2\Lambda^3}{\epsilon_o}
e^{-2\pi\frac{D}{\Lambda}}.
\end{eqnarray}
It should be noted here that the domain size $\Lambda$ is expected to depend
on the distance between assemblies.

A possible fitting of the results (Fig.~\ref{fig-5}~(a)), which sheds some
light on the origin of the attraction and also includes the $\chi-$dependence
is
\begin{eqnarray}
\Delta U_{fit}^{(d)}/N = U_{(d)}/N - P_1D^{-P_2}, \label{fitting}
\end{eqnarray}
where $d$ is the space dimensionality, $U_{2d}/N= -E_1e^{-DE_2}$ and
$U_{1d}/N=-E_1e^{-DE_2}/\sqrt{D}$, and $E_1, E_2, P_1$ and $P_2$ are fitting
parameters. Taking into account the $D$ dependence of the domain size
$\Lambda$, we can expect more complex dependence on the separation distance.

Figures~\ref{fig-5}~(c-f) shows the dependencies of all fitting parameters on
$\chi$ and compares the corresponding fitting parameters with the parameters
we computed analytically. When the system is nearly isotropic, the attraction
always includes a short--range (exponential) component and a long-range
(power law) component. The short--range term $(E_1, E_2)$ is found in the
whole range of $\chi$, suggesting correlations even for small degrees of net
incompatibility. Meanwhile the long--range term $(P_1, P_2)$ is found only
for $\chi<\chi^{(1)}$, when the system is mostly isotropic. At the point
$\chi^{(1)}$ the system undergoes a crossover transition from isotropic to
periodic phase and that affects the nature of interaction of the aggregates.
For $\chi<\chi^{(1)}$ both parameters $E_1$ and $P_1$, those determine the
strength of each component of the attraction, exponentially grow with $\chi$,
while for $\chi>\chi^{(1)},$ the long--range term diminishes and the short
range term changes the dependence on $\chi$. In the strong segregation limit,
$\chi>\chi_o$, the long--wavelength density fluctuations and therefore the
power law component of the attraction are suppressed by strong domain
correlations and the $\chi-$dependence of the exponential term changes with
respect to the $\chi<\chi_o$ limit. The multiplicative coefficient $E_1$
remains nearly constant, while the exponential decay $E_2$ starts decreasing
suggesting a growth in the range of attraction. This is due to the fact that
the domain size on the face of the interacting cylinders also grows
(Figs.~\ref{fig-4}~(a)). In the limit of large $\chi>\chi^{(1)}$, since there
no long-ranged order in 1D, the long--wavelength density fluctuations are
suppressed together with the long--range component of the attraction.

\section{Conclusions}
Compositional heterogeneities and the changes induced by nearby assemblies
via correlations and polarization, are key in the design of functional
supramolecular assemblies. Correlated domains effectively cancel the
electrostatic penalty of each other and amplify compositional segregation.
This leads to strong attractions. The cylindrical geometry results in the
breaking of the symmetry of the surface patterns, which is important in
triggering functional properties. For example, in systems with competing
interactions it can explain aggregation of co-assemblies into clusters of
different symmetries. The bare length scale of the charge surface
heterogeneity and the degree of polarization determine the symmetry of the
bundles of fibers perpendicular to their long axis and unfolds the origin of
hexagonal, square or linear (ribbons of fibers) symmetries of supramolecular
structures. These effects are of general importance in the organization of
biological assemblies. Our results not only explain the mechanism of the
attraction between both planar and curved surfaces with charged
heterogeneities, but also give insight to the understanding of electrostatic
effects in molecular biology\cite{Whitesides,SandersPRL2005,Levin}.

A recent experiment~\cite{IsraelachviliPNAS2005} shows the strongly
correlated attraction mechanism discussed here among negatively charged
surfaces (mica) coated with cationic hydrophobic surfactants (DODA). The
surfactants segregate into charged bilayer domains. This leads to a strong
attraction among coated parallel surfaces due to the growth of the bilayer
charged domains and the correlations between them and the bare mica domains
on the neighboring surface.

In summary, we have studied interaction among assemblies with surface
compositional heterogeneities. The transition from isotropic to ordered phase
leads to formation of charged domains along the cylindrical assembly. The
attraction appears as a result of correlations and polarization of charged
domains. We show that the interaction among assemblies belongs to the general
class of fluctuation--induced forces and includes long-- and short--range
term. Formation of periodic domains along the cylinder leads to the
long--range order of charged domains that suppresses long--wavelength density
fluctuations and, thus, long--range component of the attraction. Attraction
correlates and polarizes charged domains. Our results reveal the importance
of charge heterogeneities in cationic-anionic co-assemblies of complex
molecules and suggest a promising strategy for fabrication of assemblies with
predictable surface charged patterns for developing functional biomolecular
assemblies.

This work is supported by NSF grant numbers DMR-0414446 and DMR-0076097.

\newpage
\pagebreak

\begin{figure}
\caption{\label{fig-1}Snapshots of typical configurations of (a)~isotropic
phase~($\chi=1.5$) and (b)~striped state with defects~($\chi=10.5$).
$D/\sigma=3$, $R_c/\sigma=5$ and $L_z/\sigma=100$.}
\end{figure}

\begin{figure}
\caption{\label{fig-2}The dependence of (a)~normalized energy $U/\chi N$ and
(b)~dimensionless, normalized by~$N,$ heat capacity versus~$\chi$ for
different separation distances $D$.}
\end{figure}

\begin{figure}
\caption{\label{fig-crit}(a)~The density correlation function $s_{11}(q)$ for
different separation distances $D$ and (b)~the RPA critical $\chi_c$ as a
function of $D$.}
\end{figure}%

\begin{figure}
\caption{\label{fig-4} (a) Snapshots of typical configurations of surface
domains on the faced sites of separated and nearby assemblies for~$\chi=3.0,
6.0$ and $9.0$. (b, c) Dependence of $q*$ on the separation distances $D$
calculated in the simulation and numerically ($\partial s_{11}/\partial
q=0$).}
\end{figure}%

\begin{figure}
\caption{\label{fig-5}(a)~Normalized energy of interaction (points) for two
cylindrical assemblies separated by the distance $D$ for different degrees of
$\chi$. Solid lines present results of fitting of normalized energy of
interaction (Eq.~\ref{fitting}), (b)~energy of interaction calculated
according to Eq.~\ref{dE2d}, (c-f)~fitting parameters $E_1, E_2, P_1$ and
$P_2$ versus $\chi$.}
\end{figure}%

\newpage \pagebreak \clearpage

\begin{figure}
\includegraphics[width=8.7cm]{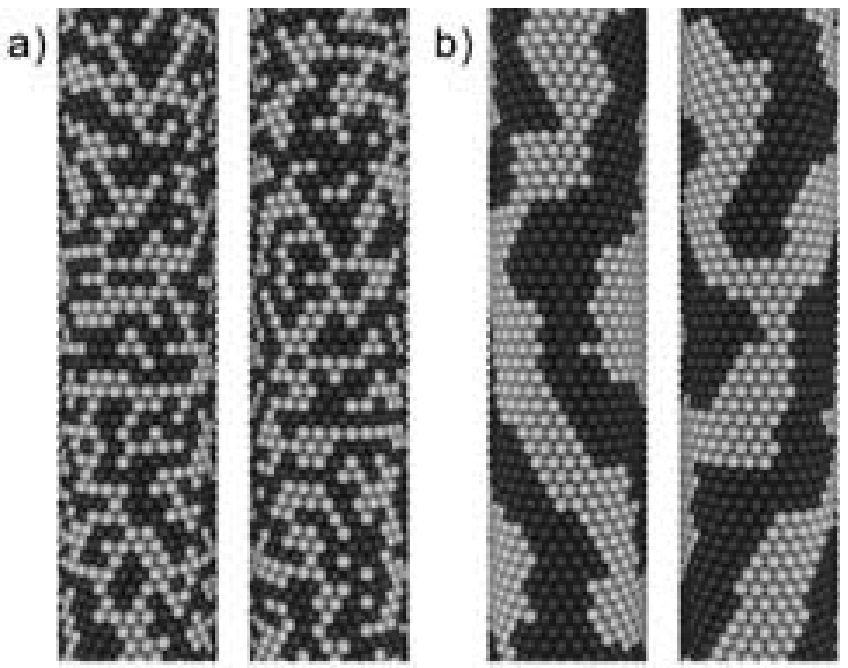}\\ Figure 1
%\caption{\label{fig-1}Snapshots of typical configurations of (a)~isotropic
%phase~($\chi=1.5$) and (b)~striped state with defects~($\chi=10.5$).
%$D/\sigma=3$, $R_c/\sigma=5$ and $L_z/\sigma=100$.}
\end{figure}
\newpage \pagebreak \clearpage

\begin{figure}
\includegraphics[width=8.4cm]{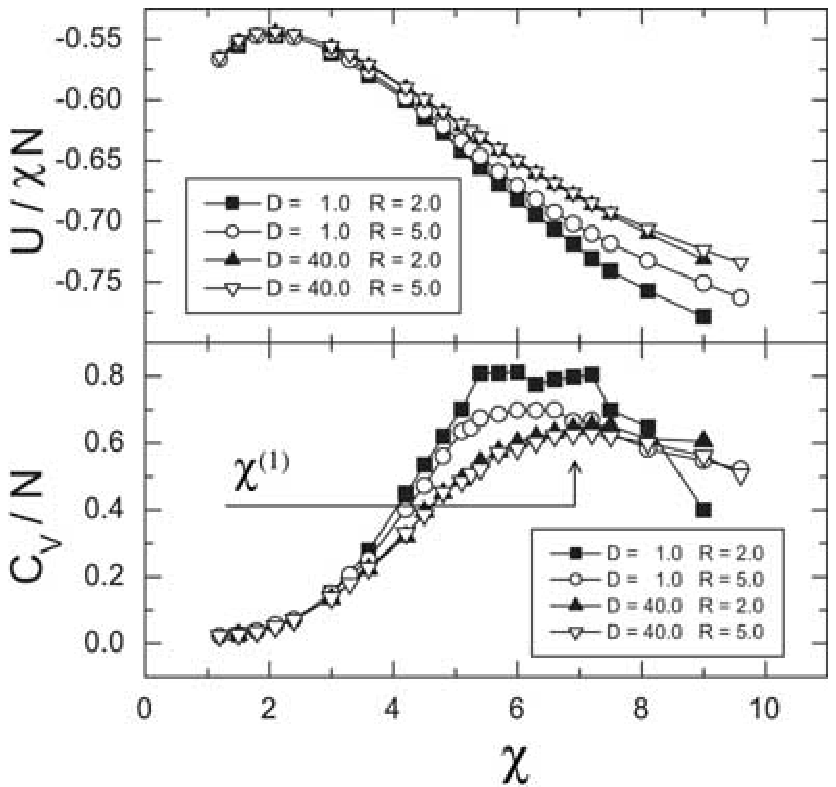}\\ Figure 2
%\caption{\label{fig-2}The dependence of (a)~normalized energy $U/\chi N$ and
%(b)~dimensionless, normalized by~$N,$ heat capacity versus~$\chi$ for
%different separation distances $D$.}
\end{figure}
\newpage \pagebreak \clearpage

\begin{figure}
\includegraphics[width=8.4cm]{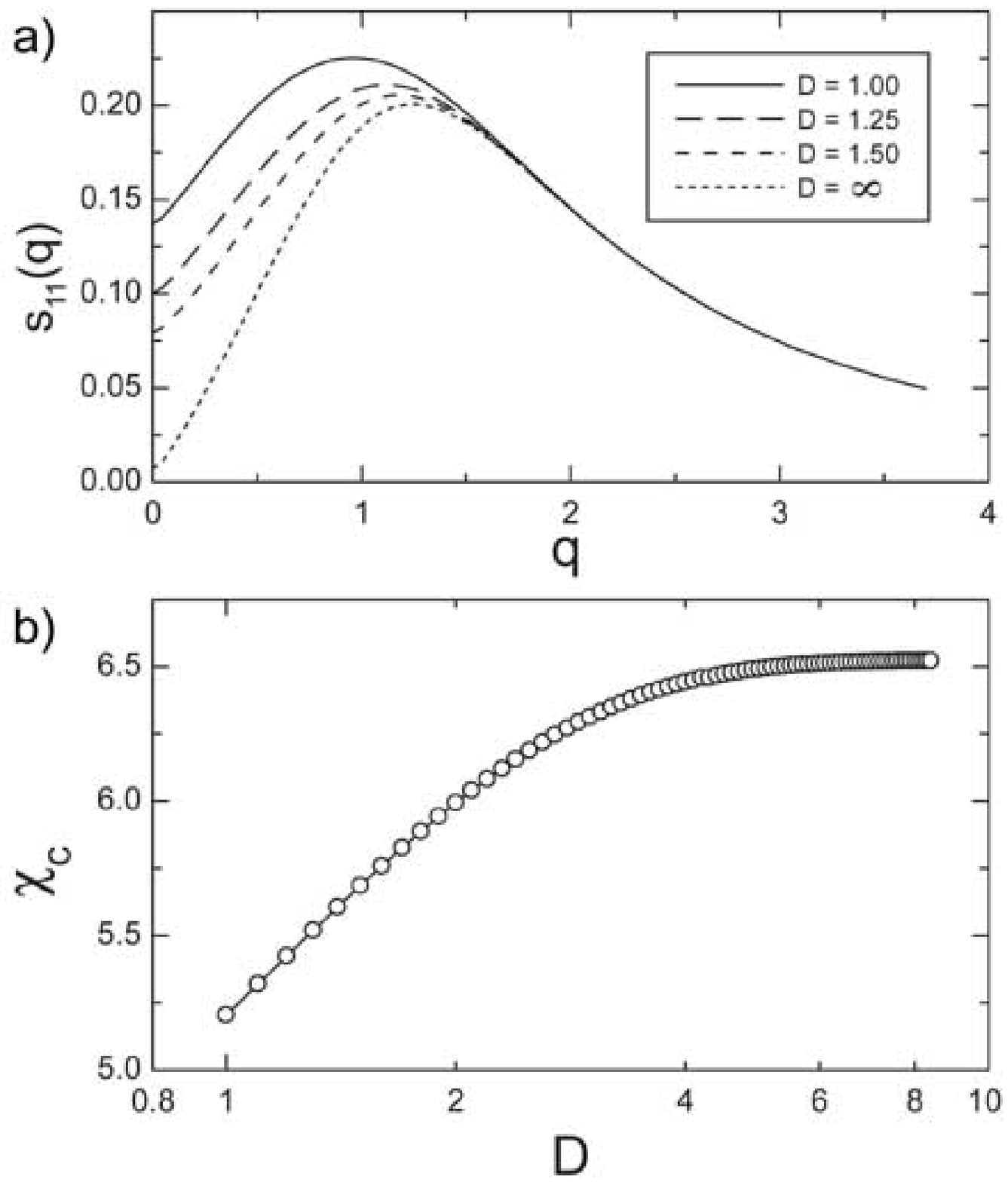}\\ Figure 3
%\caption{\label{fig-crit}(a)~The density correlation function $s_{11}(q)$ for
%different separation distances $D$ and (b)~the RPA critical $\chi_c$ as a
%function of $D$.}
\end{figure}
\newpage \pagebreak \clearpage

\begin{figure}
\includegraphics[width=8.4cm]{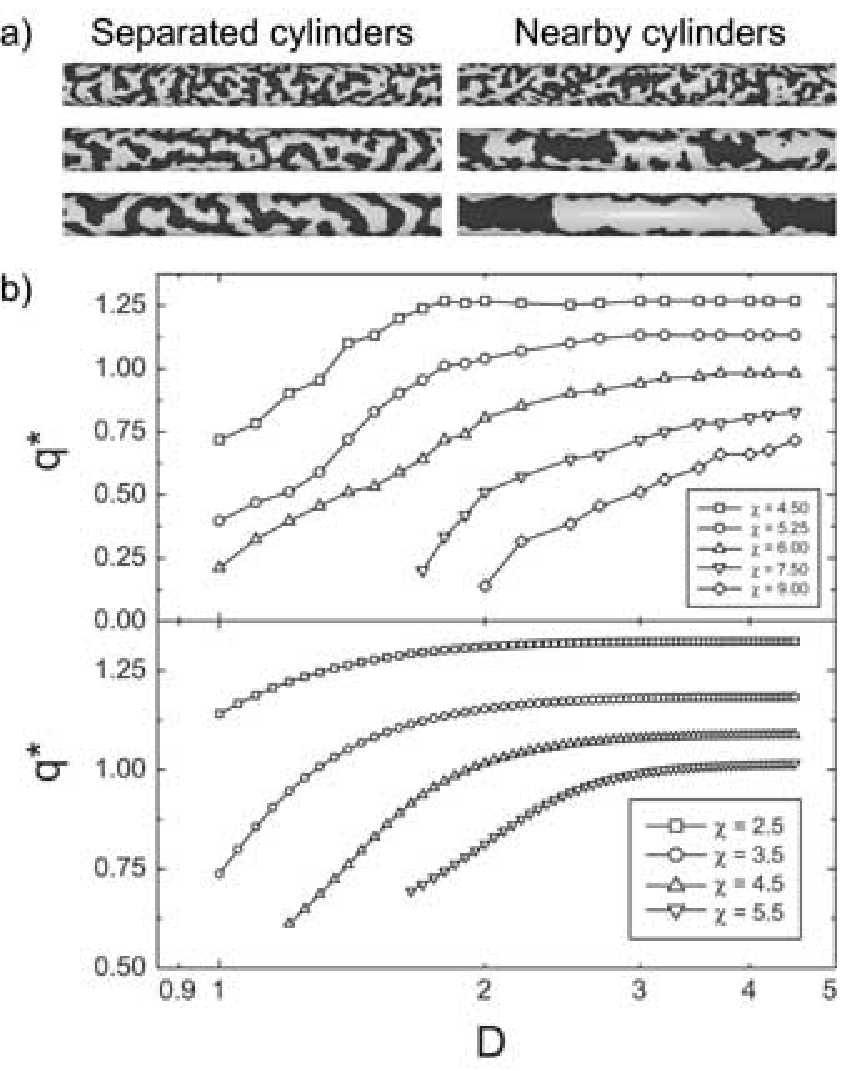}\\ Figure 4
%\caption{\label{fig-4} (a) Snapshots of typical configurations of surface
%domains on the faced sites of separated and nearby assemblies for~$\chi=3.0,
%6.0$ and $9.0$. (b, c) Dependence of $q*$ on the separation distances $D$
%calculated in the simulation and numerically ($\partial s_{11}/\partial
%q=0$).}
\end{figure}
\newpage \pagebreak \clearpage

\begin{figure*}
\includegraphics[width=17cm]{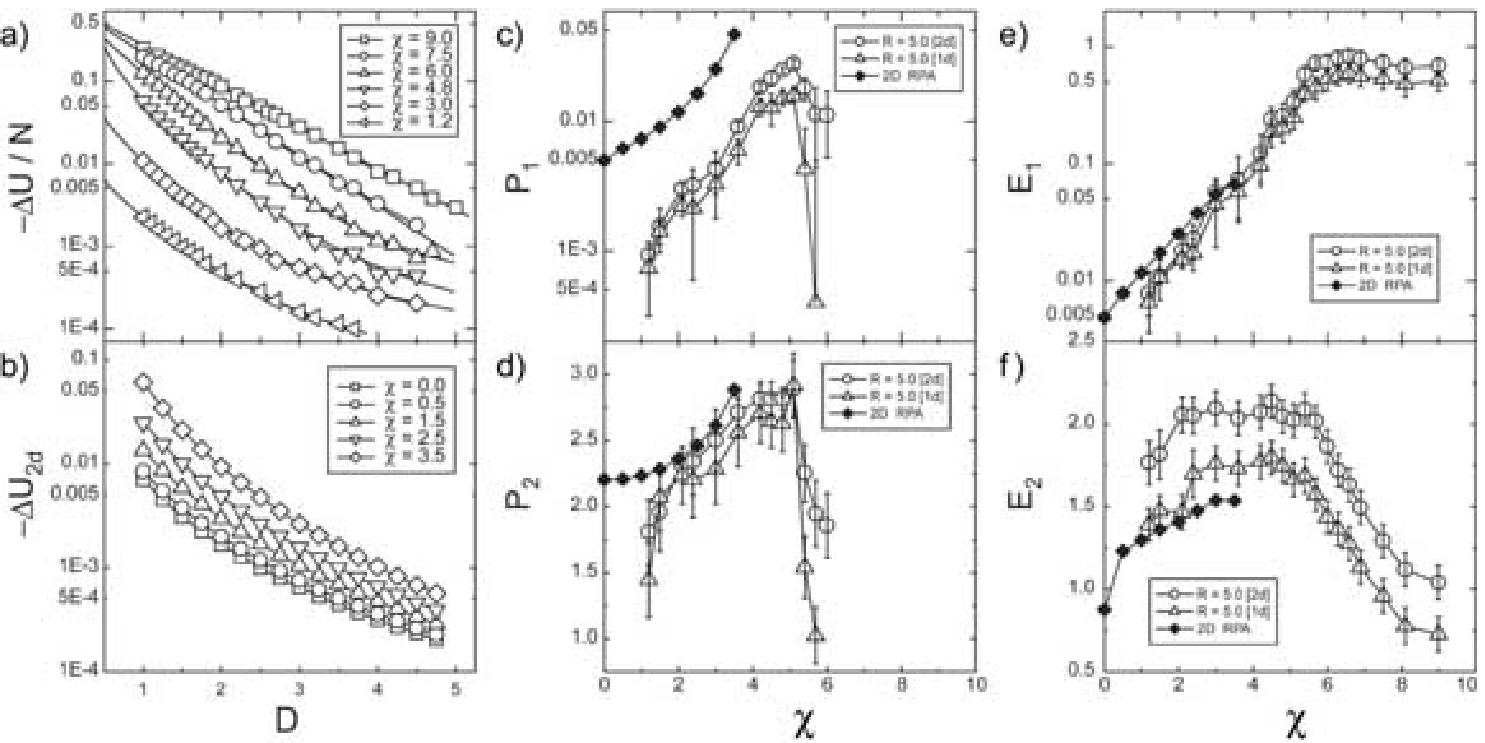}\\ Figure 5
%\caption{\label{fig-5}(a)~Normalized energy of interaction (points) for two
%cylindrical assemblies separated by the distance $D$ for different degrees of
%$\chi$. Solid lines present results of fitting of normalized energy of
%interaction (Eq.~\ref{fitting}), (b)~energy of interaction calculated
%according to Eq.~\ref{dE2d}, (c-f)~fitting parameters $E_1, E_2, P_1$ and
%$P_2$ versus $\chi$.}
\end{figure*}
\newpage \pagebreak \clearpage

\end{document}